\documentclass[preprint,showpacs,preprintnumbers,amssymb,nofootinbib]{revtex4}

\begin{document}

\title{Gravitational Optics: \\Self-phase modulation and harmonic cascades}

\author{J.T.Mendon\c{c}a}
\email{titomend@ist.utl.pt}
\affiliation{GoLP, Instituto Superior T\'{e}cnico,\\ 1049-001
Lisboa, Portugal
}%

\author{Vitor Cardoso}
\email{vcardoso@fisica.ist.utl.pt}
\affiliation{
CENTRA, Departamento de F\'{\i}sica, Instituto Superior T\'ecnico,
Av. Rovisco Pais 1, 1049-001 Lisboa, Portugal
}%

\date{\today}

\begin{abstract}

Nonlinear wave interaction of low amplitude gravitational waves in
flat space-time is considered. Analogy with optics is established.
It is shown that the flat metric space-time is equivalent to a
centro-symmetric optical medium, with no second order
susceptibility. The lowest order nonlinear effects are those due
to the third order nonlinearity and include self-phase modulation
and high harmonic generation. These processes lead to an efficient
energy dilution of the gravitational wave energy over an
increasingly large spectral range.

\end{abstract}


\maketitle
\newpage
\section{Introduction}

Gravitational radiation is a direct result of the theory of
gravitation \cite{weinberg,landau} and, even if not yet directly
observed, its existence has been indirectly inferred from the
change of orbital parameters in binary star systems \cite{taylor}.
Potential gravitational wave sources \cite{grish} can emit intense
and short bursts, such as collapsing massive binary stars or
supernova explosions, and continuous waves of much lower
amplitudes, such as stable binary stars or non-axisymmetric
spinning stars.

It is well known that the Einstein's equation governing the
gravitational field contains strong nonlinearities. The nonlinear
effects will, of course, be more relevant to the short bursts,
containing only a few cycles of radiation, and associated with the
more catastrophic events. It is also known that nonlinear
gravitational waves of the soliton type can be excited
\cite{belinskii,carr,ibanez,feinstein}. We have learned, however,
from the more common situations of nonlinear optics
\cite{bookopt}, that the excitation of solitons depends very
critically on the balance between the dispersive and the nonlinear
effects. Even if the gravitational solitons do not seem to display
such a balance and behave much more like superpositions of linear
waves \cite{boyd}, they nevertheless correspond to particular
solutions. And, with the exception of the soliton representation of
cosmological solutions \cite{belinskii2,ellis,feinstein2}, which
are pertinent to the large scale structure of the universe, it is
very unlikely that solitons will spontaneously be excited by local
gravitational wave sources.

It is therefore quite plausible in physical terms to consider the
problem of gravitational wave-packets of arbitrary shape. This will be
considered here. In order to understand their typical behavior, and to
describe the main physical pictures of their spectral evolution, we
concentrate here on the simplest possible situation: that of
propagation in a flat space-time.  We are not interested in the wave
generation phenomena, for which there exists a standard powerful
formalism \cite{isaacson} valid even in strongly curved regions, 
nor in the propagation of gravitational
waves in strongly curved regions.  Here we want to focus on regions
far from the sources, where the background is approximately flat.
This allows us to simplify considerably the problem but still
retaining some important physical features.  We will try then to
establish clear analogies with the well known concepts of nonlinear
optics. In particular, the nonlinear gravitational susceptibilities
will be established and the nonlinear wave mixing processes will be
considered. Explicit analytical solutions for the processes of
self-phase modulation and of high-order harmonic cascades will be
derived.

\section{Basic equations}

Einstein's equation for the gravitational field in the absence of
matter is given by:

\begin{equation}
R_{ik} = \Gamma_{ik,l}^l - \Gamma_{il,k}^l+
\Gamma_{ik}^l\Gamma_{lm}^m-\Gamma_{il}^m\Gamma_{km}^l = 0, \label{eq:2.1}
\end{equation}
where the comma stands for ordinary derivative, and the Christofell symbols 
are determined by the derivatives of
the metric tensor elements, according to:

\begin{equation}
\Gamma_{ik}^l = \frac{1}{2} g^{lm} \left( \frac{\partial
g_{mi}}{\partial x^k} + \frac{\partial g_{mk}}{\partial x^i} -
\frac{\partial g_{ik}}{\partial x^m} \right). \label{eq:2.2}
\end{equation}

We are considering a flat space-time, perturbed by a small
amplitude gravitational wave. This means that we can write:

\begin{equation}
g_{ij} = \eta_{ij} + h_{ij}, \label{e:2.3} \end{equation} where
$|h_{ij}| \ll 1$, and:

\begin{equation}
\eta_{00} = 1 \quad , \quad \eta_{ii} = -1 \quad (i = 1, 2, 3)
\quad , \quad  \eta_{ij} = 0 \quad (i \neq j). \label{eq:2.4}
\end{equation}
In this case, we can write:

\begin{equation}
R_{ik} = R_{ik}^L + R_{ik}^{NL}, \label{eq:2.5} \end{equation}
where the linear term is:

\begin{equation}
R_{ik}^L = \frac{1}{2} \eta^{lm} \left( \frac{\partial^2
h_{mi}}{\partial x^l \partial x^k} + \frac{\partial^2
h_{mk}}{\partial x^l \partial x^i} -\frac{\partial^2
h_{ik}}{\partial x^l \partial x^m} - \frac{\partial^2
h_{ml}}{\partial x^k \partial x^i} \right). \label{eq:2.6}
\end{equation}

The nonlinear term $R_{ik}^{LN}$ contains second, third and higher
order nonlinearities. This means that, in principle, we can have
three, four and higher wave mixing processes. Let us then write:

\begin{equation}
R_{ik}^{LN} = R_{ik}^{(2)} + R_{ik}^{(3)} + ...,
\label{eq:2.7} \end{equation} where we have, for the lowest order
term:

\begin{equation}
R_{ik}^{(2)} = \frac{h^{lm}}{\eta^{lm}} R_{ik}^L +
\frac{\eta^{lp}}{h^{lp}} R_{ik}^{(3)} + \frac{1}{2} \left[
\frac{\partial h^{lm}}{\partial x^l} \left( \frac{\partial
h_{mi}}{\partial x^k} + \frac{\partial h_{mk}}{\partial x^i} -
\frac{\partial h_{ik}}{\partial x^m} \right) -\frac{\partial
h^{lm}}{\partial x^k}\frac{\partial h_{ml}}{\partial x^i} \right].
\label{eq:2.8} \end{equation}
The third order nonlinearities are
contained in:

\begin{eqnarray}
R_{ik}^{(3)} = - \frac{1}{4} \eta^{nm} h^{lp} \left[ \left(
\frac{\partial h_{mi}}{\partial x^l} + \frac{\partial
h^{ml}}{\partial x^i} -\frac{\partial h_{il}}{\partial x^m}
\right) \left( \frac{\partial h_{pk}}{\partial x^n}
+\frac{\partial h_{pn}}{\partial x^k} - \frac{\partial
h_{kn}}{\partial x^p} \right) \right. \nonumber \\ - \left. \left(
\frac{\partial h_{pi}}{\partial x^k} +\frac{\partial
h_{pk}}{\partial x^i} - \frac{\partial h_{ik}}{\partial x^p}
\right) \frac{\partial h_{np}}{\partial x^l} \right].
\label{eq:2.9} \end{eqnarray}
We could go on to higher orders but as we assume $|h_{ij}| \ll 1$, these 
terms become less important. Furthermore, stopping at third order
suffices, as we will show, to uncover some truly non-linear and interesting
aspects of gravitation.

It is well known \cite{landau} that
the linear term (\ref{eq:2.6}) can simply be written as:

\begin{equation}
R_{ik}^L \simeq - \frac{1}{2}\partial^{\mu}\partial_{\mu}  h_{ik}, \label{eq:2.11}
\end{equation} 
where the d'Alembert operator is:

\begin{equation}
\partial^{\mu}\partial_{\mu} = \frac{\partial^2}{\partial x^j \partial x_j} = \eta^{jn}
\frac{\partial^2}{\partial x^j \partial x^n}. \label{eq:2.12}
\end{equation}
This means that we can write the nonlinear wave equation as:

\begin{equation}
\partial^{\mu}\partial_{\mu} h_{ik} = 2 \left[ R_{ik}^{(2)} + R_{ik}^{(3)} +
R_{ik}^{(4)} \right]. \label{eq:2.13} \end{equation}

\section{Nonlinear wave coupling}

Equation (\ref{eq:2.13}) predicts the possibility of several types
of nonlinear gravitational wave coupling. This can be studied by
assuming that several waves of the type:

\begin{equation}
h_{ij} (n) = \epsilon_{ij} (n) a (n) \exp[i q_k (n) x^k],
\label{eq:3.1} \end{equation} 
coexist in the same region of flat
space-time. Here $a (n)$ are slowly varying amplitudes,
$\epsilon_{ij} (n)$ are unit polarization tensors such that
$\epsilon_{ij}^* \epsilon_{ij}=1$, and $q_k (n)$ are the
four-wavevectors associated with the interacting waves
$n=1,2,3...$

The total wave field, for real waves, will be given by:

\begin{equation}
h_{ij} = \sum_n h_{ij} (n) + c.c. \label{eq:3.2} \end{equation} 
In order to determine the evolution of the amplitudes $a (n)$, we can
construct envelope equation, by noting that:

\begin{equation}
\partial^{\mu}\partial_{\mu} h_{ij} (n) \simeq \left( - q_k q^k + i q_k
\frac{\partial}{\partial x_k} \right) h_{ij} (n). \label{eq:3.3}
\end{equation}
Assuming that the waves still obey the linear dispersion relation,
valid in the limit $R_{ij}^{NL} \rightarrow 0$:

\begin{equation}
q_k (n) q^k (n) = 0, \label{eq:3.4} \end{equation} 
we obtain, for three wave mixing:

\begin{equation}
i q_k (1) \frac{\partial}{\partial x_k} a (1) = v (1) a (2) a (3)
\exp(i \Delta q_k x^k), \label{eq:3.5} \end{equation} 
where:

\begin{equation}
\Delta q_k = q_k (2) + q_k (3) - q_k (1). \label{eq:3.6}
\end{equation}

The nonlinear coupling coefficient $v (1)$ is determined by
$R_{ij}^{(2)}$, equation(\ref{eq:2.8}). It can be verified that $v
(1) = 0$, which means that the three wave coupling is forbidden.
We can use the hard way, by calculating explicitly $R_{ij}^{(2)}$,
or the easy way, by noting that flat space-time is a
centro-symmetric medium and therefore, like in optics, it has no
second order susceptibility. We then proceed to the next order of
nonlinearity, and write the envelope equation for the four wave
mixing process:

\begin{equation}
i q_k (1) \frac{\partial}{\partial x_k} a (1) = w (1) a^* (2) a
(3) a(4) \exp(i \Delta q_k x^k), \label{eq:3.7} \end{equation}
where now:

\begin{equation}
\Delta q_k = q_k (3) + q_k (4) - q_k (1) - q_k (2), \label{eq:3.8}
\end{equation} 
and the coupling coefficient $w (1)$ is determined
by $R_{ij}^{(3)}$. Similar equations given the other three wave
amplitudes. In order to derive explicit results, we assume wave
propagation along a given direction $x^1 = x$. We have: $i q_k x^k
= i q_0 x^0 + i q_1 x^1$. We will use: $q_1 = q$, $q_0 = -
\Omega/c$, and $x^0 = c t$, which then leads to:

\begin{equation}
\exp(i q_k x^k) = \exp(i q x - i \Omega t). \label{eq:3.9}
\end{equation}
The linear dispersion relation (\ref{eq:3.4}) is:

\begin{equation}
q_k x^k = \frac{\Omega^2}{c^2} - q^2 = 0. \label{eq:3.10}
\end{equation}
This shows that flat space-time is non-dispersive. This means
that, for $\Omega(2) + \Omega (1) = \Omega (1) + \Omega (2)$, we
can guarantee a perfect phase matching:

\begin{equation} \Delta q_k = 0. \label{eq:3.11} \end{equation}
We will also have:

\begin{equation}
i q_k \frac{\partial}{\partial x_k} = i q_k \eta^{kj}
\frac{\partial}{\partial x^j} = i q \left(
\frac{\partial}{\partial x} +  \frac{1}{c}
\frac{\partial}{\partial t} \right). \label{eq:3.12}
\end{equation}
We can now introduce a variable $z = x - c t$, such that:

\begin{equation}
i q (1) \frac{\partial}{\partial z} a (1) = w (1) a^* (2) a (3) a
(4). \label{eq:3.13} \end{equation} Let us use equation
(\ref{eq:2.9}) to calculate. The result is:

\begin{eqnarray}
w (1) = \frac{1}{4} \epsilon^{ik*} \eta^{nm} \epsilon^{lp*} (2)
\left\{  \left[ q_l (3) \epsilon_{mi} (3) + q_i (3) \epsilon_{ml}
(3) - q_m (3) \epsilon_{il} (3) \right] \right. \cdot \nonumber \\
\left[ q_n (4) \epsilon_{pk} (4) + q_k (4) \epsilon_{pn} (4) - q_p
(4)\epsilon_{kn} (4) \right] - q_l (4) \epsilon_{np} (4) \nonumber
\\ \left. \left[ q_k (3) \epsilon_{pi} (3) + q_i (3)
\epsilon_{pk} (3) - q_k (3) \epsilon_{ik} (3) \right] \right\} +
permutation (2,3,4). \label{eq:3.14} \end{eqnarray} 
This is a very
complicated expression, where the last term represents two terms
formally identical to the first one, and obtained by permutation
of the indices $2, 3$ and $4$. It is important to notice that, in
order of magnitude, we have $w (1) \simeq \Omega^2 / c^2$.

\section{Self-phase modulation}

An important particular case is obtained when the modes $n = 1, 2,
3, 4$ coincide. The mode coupling equation (\ref{eq:3.7}) is then
reduced to:

\begin{equation}
i q \frac{\partial}{\partial z} a = w |a |^2 a. \label{eq:4.1}
\end{equation}
It is instructive to compare it with the envelope equation for an
optical pulse (for instance a laser pulse), with central frequency
$\omega_0$, propagating in a nonlinear medium with group velocity
$v_0$. Its electric field amplitude is described by the equation
\cite{bookopt}:

\begin{equation}
\left( \frac{\partial}{\partial x} + \frac{1}{v_0}
\frac{\partial}{\partial t} \right) E_0 = i \omega_0 \alpha |E_0|^2
E_0. \label{eq:4.2} \end{equation} 
where $E_0$ is the electric
field amplitude and $\alpha$ is the nonlinear coupling
coefficient:

\begin{equation}
\alpha = \frac{\omega_0}{k_0 c^2} \chi^{(2)}. \label{eq:4.3}
\end{equation}
Notice that equation can be written in a similar form:

\begin{equation}
\left( \frac{\partial}{\partial x} + \frac{1}{c}
\frac{\partial}{\partial t} \right) a = - i \frac{w}{q} |a|^2 a.
\label{eq:4.31} \end{equation}

Comparing with equation (\ref{eq:4.2}) we conclude that the
equivalent nonlinear susceptibility of the flat space-time, which
would be the gravitational wave version of the electromagnetic
susceptibility $\chi^{(2)}$ is equal one, whereas in optical
materials the typical value is $\chi^{(2)} \sim 10^{-16}$. This
means that he gravitational field is extremely nonlinear and that
we expect to observe the same nonlinear effects as in optics at a
much lower wave amplitudes. In particular, we can follow the usual
optical approach and obtain a wavepacket solution of the form:

\begin{equation}
a (x, t) = a (z) \exp \left[ i \phi (z, t) \right], \label{eq:4.4}
\end{equation} where the nonlinear phase is:

\begin{equation}
\phi (z, t) = \phi_0 - \frac{w}{q} |a (z)|^2 t. \label{eq:4.5}
\end{equation}
This means that a gravitational wavepacket, even of very small
amplitude, will suffer a self-phase modulation such that its
central frequency $\Omega$ will not remain fixed but will change
along propagation, according to:

\begin{equation}
\Omega = \Omega (0) - \frac{\partial \phi}{\partial t} = \Omega
(0) - \frac{w}{qc} t \frac{\partial}{\partial z} |a (z)|^2.
\label{eq:4.6} \end{equation} 
For instance, if we have a gaussian
gravitational wave pulse of the form:

\begin{equation}
|a (z)|^2 = a_0^2 \exp \left( - \frac{z^2}{z_0^2} \right),
\label{eq:4.7} \end{equation} we will obtain a gravitational wave
frequency shift $\Delta \Omega = \Omega - \Omega(0)$:

\begin{equation}
\Delta \Omega = \frac{w}{q c} \frac{2 z}{z_0^2} |a (z)|^2 t.
\label{eq:4.8} \end{equation}

It contrast with what usually occurs in nonlinear optics, this
frequency shift will be positive at the pulse front ($z > 0$), and
negative at the pulse rear ($z < 0$). We also see that, for very
large propagation times $t$ this shift can be quite significant,
$\Delta \Omega \gg \Omega (0)$, even for weak gravitational
perturbations, $a \ll 1$, due to the already noticed strong
nonlinearities.
To be quantitative, let's see what conditions have to be fulfilled in
order to have the same effect as in nonlinear optics, i.e., to have
the same frequency shift $\frac{ \Delta \Omega}{\Omega}_{\rm
opt}=\frac{ \Delta \Omega}{\Omega}_{\rm grav}$.  In order to satisfy
this condition we have to have

\begin{equation}
 |a (z)|^2 \sim |E_0|^2 \chi^{(2)}_{opt}\frac{d_{opt}}{d_{grav}},
\label{eq:4.9} \end{equation}
where $d_{opt}$ is a length travelled by light in usual experiments,
and $d_{grav}$ is the length travelled by a gravitational wave until
it reaches the gravitational wave detector.  As a typical example, we
have $d_{opt} \sim 10^{-3} {\rm m}\,,\,|E_0|^2 \chi^{(2)}_{opt}\sim
10^{-2}$.  For these values, we get
\begin{equation}
 |a (z)| \sim \frac{6 \times 10^{-13}}{\left[d_{grav}(Kpc)\right]^{1/2}}\,,
\label{eq:4.10} \end{equation}
where the distance is now measured in $Kpc$.
For a gravitational wave produced by any reasonable
astrophysical event, our best expectations \cite{isaacson} give $ a \sim 10^{-21}$, 
as the wave arrives on Earth.
But, along its path from the radiation source
to the Earth, its amplitude will remain significantly higher over
large distances. It seems therefore  very likely that such an effect
will indeed take place.

\section{Harmonic cascade}

Another possibility of nonlinear gravitational wave coupling is
the generation of a large spectrum of harmonics of the initial
frequency $\Omega (0)$. Here, in principle, the sum over $n$ will
extend to infinity:

\begin{equation}
h_{ij} = \sum_{n = - \infty}^{\infty} h_{ij} (n). \label{eq:5.1}
\end{equation}
We will also have:

\begin{equation}
q_k (n) = n q_k.  \label{eq:5.2} \end{equation} Four wave mixing
between these different harmonics will occur, where a perfect
phase-matching condition can be verified:

\begin{equation}
\Delta q_k = \left[ q_k (2) + q_k (n - 1) \right] - \left[ q_k (1)
+ q_k (n) \right] = 0. \label{eq:5.3} \end{equation}

For the present case, an envelope equation for the amplitudes of
the different harmonics can be derived, in the same way as
equation (\ref{eq:3.7}). Assuming that the fundamental and the
first harmonics are dominant, $|h_{ij} (n = 1, 2) \gg h_{ij} (n
\neq 1, 2)$, we can retain only two of the mixing terms and get,
for propagation along one given direction $x$:

\begin{equation}
i q(n) \left( \frac{\partial}{\partial x} + \frac{1}{c}
\frac{\partial}{\partial t} \right) a (n) = w (n) \left[ a^* (1) a
(2) a (n - 1) + a (1) a^* (2) a (n + 1) \right]. \label{eq:5.4}
\end{equation} 
where $w (n)$ is determined by an expression
similar to equation (\ref{eq:3.14}). Let us now define:

\begin{equation}
a^* (1) a (2) = I e^{i \delta}, \label{eq:5.5} \end{equation}
where $I$ and $\delta$ are real. In order to get order of
magnitude analytical solutions, let us assume that $w (n) \propto
q (n)$, and let us also neglect the amplitude variations of the
dominant modes, which means that the intensity parameter $I$ is
approximately constant. We are then led to:

\begin{equation}
\frac{\partial}{\partial z} a (n) = i w \left[ e^{i \delta} a (n -
1) + e^{- i \delta} a (n + 1) \right], \label{eq:5.51}
\end{equation} with:

\begin{equation}
w = - \frac{w (n)}{q (n)} I. \label{eq:5.6} \end{equation} 
It is now convenient to introduce new amplitude variables $b (n)$, such
that:

\begin{equation}
a (n) = (- 1)^{n/2} e^{i n \delta} b (n), \label{eq:5.7}
\end{equation} 
and we get:

\begin{equation}
\frac{\partial}{\partial z} = w \left[ b (n - 1) - b (n + 1)
\right]. \label{eq:5.8} \end{equation} 
Using $\tau = w z$, we can
reduce this equation to the well known recurrence relation for
Bessel functions:

\begin{equation}
\frac{\partial}{\partial \tau} b (n) = b (n - 1) - b (n + 1).
\label{eq:5.9} \end{equation}

An adequate solution will then be given in terms of Bessel
functions of the first kind $J_n (\tau)$, such as:

\begin{equation}
b (n, \tau) = A J_n (\tau) + A' J_{n-1} (\tau), \label{eq:5.10}
\end{equation} where the two constants $A$ an $A'$ are determined
by the initial conditions:

\begin{equation}
A = a(n=1, \tau =0) \quad , \quad A' = a (n=2, \tau=0),
\label{eq:5.11} \end{equation} 
and all the other harmonic wave
amplitudes as assumed to be initially equal to zero: $a (n \neq 1,
2; \tau=0)$. Notice that this solution satisfies energy
conservation, in the sense that:

\begin{equation}
\sum_{n = - \infty}^{\infty} | b (n, \tau) |^2 = const.
\label{eq:5.12} \end{equation} 
It is then an adequate, if if
particular, solution to the problem of harmonic generation of low
amplitude gravitational waves propagating in a flat space-time.

\section{Conclusions}

Nonlinear wave coupling of gravitational waves was considered in
this work. The problem of low amplitude waves in flat space-time
was examined, in order to extract the main physical consequences
with the simplest possible formal complexity and to establish
clear connections with the well known effects occurring in
nonlinear optics.

In particular, the gravitational equivalent to the optical second
and third order susceptibility was derived. It was shown that the
second order gravitational susceptibility is equal to zero and,
consequently, the process of three nonlinear wave coupling is
forbidden. This is not surprising because the flat space-time
background considered here can be seen as a centro-symmetric
material medium. This property will eventually disappear in curved
space-time.

In contrast, the gravitational third order susceptibility was
shown to be equal to one, thus meaning that the flat space-time is
a strongly nonlinear medium. As in optics, four wave mixing
processes can then be considered. But, unlike the case of optical
phenomena, these wave mixing processes can become relevant even
for very low wave amplitudes.  The particularly important cases of
generation of harmonic cascades and of self phase modulation were
considered, and explicit analytical solutions were derived. They
show that a significant spectral broadening and spectral energy
dilution can then take place, specially in the vicinity of
radiation sources where the wave amplitudes are non-negligible.
Such energy dilution is not considered in the usual estimates for
gravitational wave detection \cite{grish,novikov} and could
eventually lower the prospects for direct observation of these
waves near the Earth.

Finally, it should be noticed that self phase modulation occurring
in optics can be described as a particular example of photon
acceleration \cite{luis,bookmend}. In analogy, the present result
showing the existence of self phase modulation of gravitational
wave pulses could also be interpreted as graviton acceleration.
This aspect will be explored in a future work.

\end{document}